\documentclass[letter]{aa}  

\bibpunct{(}{)}{;}{a}{}{,}
\usepackage{graphicx}
\usepackage{txfonts}
\begin{document}

   \title{Very long baseline interferometry observation of the triple AGN candidate J0849$+$1114}
   \titlerunning{VLBI observation of the triple AGN candidate J0849$+$1114}

   \author{K. \'E. Gab\'anyi
          \inst{1,} \inst{2}
          \and
          S. Frey\inst{2}
          \and S. Satyapal \inst{3}
	\and A. Constantin \inst{4}
\and R. W. Pfeifle\inst{3}
          }

   \institute{MTA-ELTE Extragalactic Astrophysics Research Group, ELTE TTK,
              P\'azm\'any P\'eter s\'et\'any 1/A, 1117 Budapest, Hungary\\
              \email{krisztina.g@gmail.com}
         \and
             Konkoly Observatory, MTA Research Centre for Astronomy and Earth Sciences, Konkoly Thege Mikl\'os \'ut 15-17 1121 Budapest, Hungary
	\and 
	Department of Physics and Astronomy, George Mason University, 4400 University Drive, MSN 3F3, Fairfax, VA 22030, USA
	\and
	Department of Physics and Astronomy, James Madison University, Harrisonburg VA 22807, USA
               }

   \date{Received , 2019; accepted , 2019}

 
  \abstract
   {In the hierarchical structure formation model, galaxies grow through various merging events. Numerical simulations indicate that the mergers can enhance the activity of the central supermassive black holes in the galaxies.} 
{\cite{Pfeifle_new} identified a system of three interacting galaxies, J0849$+$1114, and provided multiwavelength evidence of all three galaxies containing active galactic nuclei. The system has substantial radio emission, and with high-resolution radio interferometric observation we aimed to investigate its origin, whether it is related to star formation or to one or more of the active galactic nuclei in the system. }
   {We performed high-resolution continuum observation of J0849$+$1114 with the European Very Long Baseline Interferometry Network at $1.7$\,GHz.}
   {We detected one compact radio emitting source at the position of the easternmost nucleus. Its high brightness temperature and radio power indicate that the radio emission originates from a radio-emitting active galactic nucleus. Additionally, we found that significant amount of flux density is contained in $\sim 100$ milliarcsec-scale feature related to the active nucleus.}
   {}


   \maketitle
%

\section{Introduction}

According to currently accepted cosmological and structure formation model, galaxies grow through frequent mergers \cite[e.g.,][]{galaxy_grow}. During these events, the supermassive black holes (SMBHs) residing in the merging galaxies shrink to the central region of $\sim 1$\,kpc losing energy by dynamical friction. There, a bound SMBH pair may form, which eventually will be able to emit gravitational waves before their final coalescence \citep{Begelman1980}. Detection of an SMBH is relatively straightforward if it is actively accreting matter from its surrounding as in an active galactic nucleus (AGN). According to numerical simulations \citep[e.g.][]{Capelo2015,Blecha2018}, the simultaneous activity in the nuclei of merging galaxies is expected at a separation of $\lesssim 10$\,kpc. 

Despite various efforts, efficient observational selection of multiple AGN candidates has so far been elusive. \cite{Satyapal2017} proposed a new selection method to find possible dual AGN. They selected galaxies from the Sloan Digital Sky Survey (SDSS) experiencing interactions or mergers according to their optical images. This sample is cross-correlated with the infrared AllWISE catalog\footnote{\url {http://wise2.ipac.caltech.edu/docs/release/allwise/}} \citep{allwise} to identify those sources where AGN can be present. They employ a colour cut using the two shortest wavelength observing bands, W1$-$W2$>0.5$ (where W1 and W2 are the $3.4 \mu$m and $4.6\mu$m observing bands, respectively) to ensure the selection of AGN candidates, a colour-cut that has been shown to be the most effective at finding AGNs in late stage mergers based on hydrodynamic simulations \citep{Blecha2018}. They conducted X-ray follow-up observations of selected sources from this sample, where the separation of the assumed AGN can be resolved spatially by the {\it Chandra} X-ray satellite. In their pilot study, four sources from the observed six showed two distinct X-ray components indicating that they possibly host dual AGNs at separations $< 10$\,kpc. In a follow-up study, \cite{Pfeifle2019} found that $8$ out of $15$ merging galaxy systems had multiple nuclear X-ray sources suggestive of dual AGN. 

\begin{figure}
   \centering
    \includegraphics[width=\columnwidth, bb=140 25 780 670, clip=]{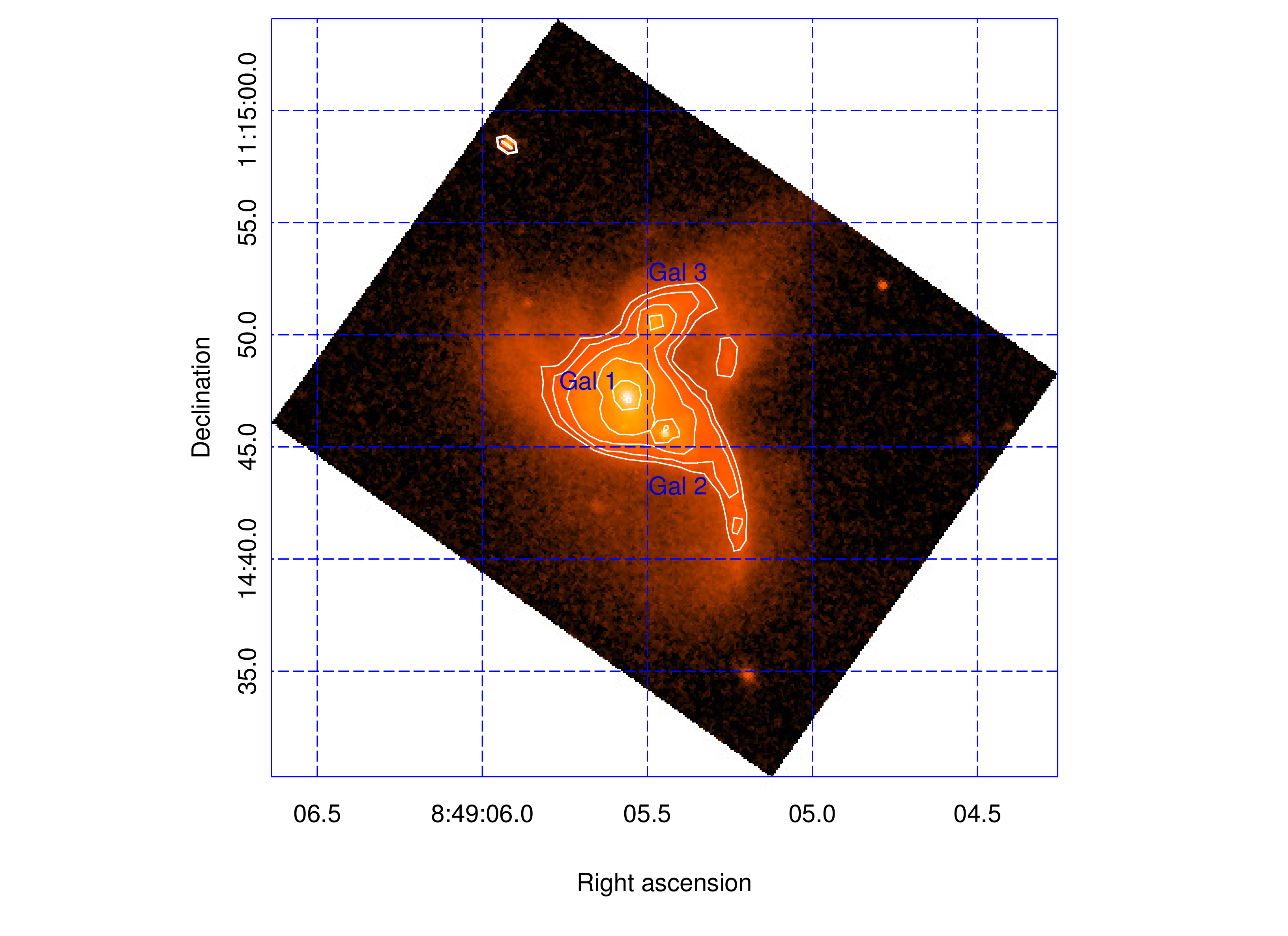}
      \caption{Archival {\it Hubble Space Telescope} WFC3 F105W image of J0849$+$1114 from \cite{Pfeifle_new}, the three interacting galaxies are marked. 
              }
         \label{fig:hst}
   \end{figure}

One of those sources, J0849$+$1114 is a system of three interacting galaxies within $\sim 5\arcsec$ projected separation (Fig. \ref{fig:hst}), all of them exhibiting nuclear X-ray sources with $2-10$\,keV luminosities in excess of the expected star formation and showing optical line ratios consistent with AGN photoionization processes \citep{Pfeifle_new,Pfeifle2019}. In the twelfth data release of the Sloan Digital Sky Survey \citep[SDSS DR12, ][]{sdss} Galaxy 1 is listed as SDSS J084905.51$+$111447.2 and Galaxy 3 as SDSS J084905.43$+$111450.9. 

Recently, \cite{nature_cikk} also reported on the existence of three AGN in J0849$+$1114. According to the authors, all three stellar nuclei can be classified as Seyfert 2 galaxies. They detected radio emission features at $9$\,GHz with the Karl G. Jansky Very Large Array (VLA) at the positions of Galaxy 1 and Galaxy 3 (Fig. \ref{fig:hst}).

Here, we report on the results of the milliarcsec (mas) scale resolution very long baseline interferometry (VLBI) observation of J0849$+$1114 performed by the European VLBI Network (EVN). In the following, we define the radio spectral index, $\alpha$ as $S\propto \nu^{\alpha}$, where $S$ is the flux density and $\nu$ is the observing frequency. We assume a flat $\Lambda$CDM cosmological model with $H_0=70\mathrm{\,km\,s}^{-1}\mathrm{\,Mpc}^{-1}, \Omega_\mathrm{m}=0.27$, and $\Omega_\Lambda=0.73$. At the redshift of J0849$+$1114, $z=0.077$, $1$\,mas angular separation corresponds to $1.46$\,pc projected linear distance \citep{wright}.

%

\section{Observation and data reduction}

\begin{table*}
\caption{EVN radio telescopes that participated in the observation.}             
\label{table:obs}      
\centering                          
\begin{tabular}{c c c}        
\hline\hline                 
Antenna name & Diameter (m) & Observing time \\    
\hline                        
Jodrell Bank Mk2 (United Kingdom) & $25$ & 20:00$-$22:05 \\      
Westerbork Synthesis Radio Telescope (the Netherlands)\tablefootmark{a} & $25$ &  20:00$-$22:05 \\
Effelsberg (Germany) & $100$ & 20:00$-$20:35 \\
Medicina (Italy) & $32$ & 20:00$-$21:40 \\
Onsala (Sweden) & $25$ & 20:00$-$22:05 \\ 
Tianma (China) & $65$ & 20:00$-$22:05 \\
Toru\'n (Poland) & $32$ & 20:00$-$22:05 \\
Hartebeesthoek (South Africa) & $26$ & 20:00$-$22:05 \\
Irbene (Latvia) & $32$ & 20:00$-$22:05 \\
Sardinia (Italy) & $65$ & 20:00$-$22:05 \\
\hline                                   
\end{tabular}
\tablefoot{
\tablefoottext{a}{One antenna was used.}
}
\end{table*}


An exploratory EVN observation of J0849$+$1114 was conducted on 2019 Jan 22 (project code: RS09b) at $\nu=1.7$\,GHz. The names, diameters and observing times of the participating antennas are given in Table \ref{table:obs}. 
At the end, data from all antennas except Medicina and Irbene were used to create the final map of the source. Eight intermediate frequency channels (IFs) each with $16$\,MHz bandwidth were used in both left and right circular polarization. Each IF was divided to $32$ spectral channels. The observation was carried out in e-VLBI mode \citep{evlbi}. The data from the antenna sites were transferred through optical fiber connection to the Joint Institute for VLBI ERIC, Dwingeloo (in the Netherlands), where they were correlated with an integration time of $2$\,s.

The observation lasted for $\sim 2$\,h, and the on-source integration time was $55$\,min. The observation was conducted in phase-reference mode \citep{phase-ref}, using J0851$+$0845 as the phase reference calibrator (target separation: $2\fdg 56$). An additional phase-reference candidate source, J0850$+$1108 (target separation: $0\fdg 42$) was also included in the observation. For the target source, we used the coordinates from the Faint Images of the Radio Sky at Twenty-centimeters survey \citep[FIRST, ][]{first_cat} as phase centre, right ascension $08^\mathrm{h} 49^\mathrm{m} 05\fs 51$, and declination $11\degr 14\arcmin 48\arcsec$. 

For calibration, we followed standard procedures \citep{data_red} using the NRAO Astronomical Image Processing System  ({\sc aips}, \citealt{aips}). The amplitudes of the interferometric visibilities were calibrated using the antenna gain curves and the system temperatures measured at the telescope sites.\footnote{Only nominal system temperature values were available for the Jodrell Bank Mk2 and the Toru\'n antennas.} Fringe-fitting was performed for the phase calibrator. The fringe-fitted data of the phase calibrator source were imaged using the {\sc difmap} software package \citep{difmap} following standard hybrid mapping procedure, which included several cycles of {\sc clean}ing \citep{clean}, phase-only self-calibration and finally amplitude self-calibration. The antenna-based gain correction factors determined in {\sc difmap} were applied to the data in {\sc aips}\footnote{The flux density of the phase calibrator source before and after applying the gain scales  was 124 mJy and 108 mJy, respectively.}. A new fringe-fit of the calibrator was performed, now taking into account the {\sc clean} model components describing its brightness distribution, to reduce the small residual phase errors due to its structure. The obtained solutions were applied to the target source J0849$+$1114 and to the candidate secondary phase reference source, J0850$+$1108. Then both of them were imaged in {\sc difmap}. 

\begin{figure}
   \centering
    \includegraphics[width=\columnwidth, bb=65 178 520 620, clip=, angle=-90]{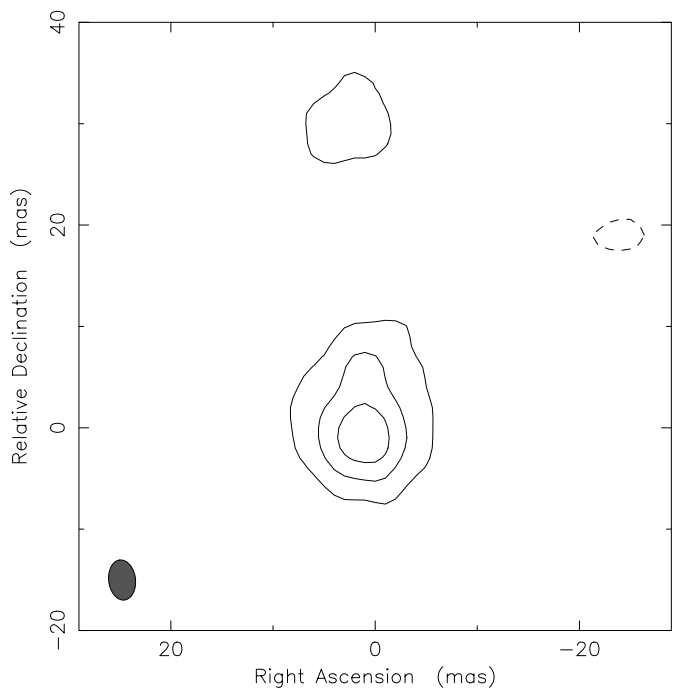}
      \caption{$1.7$-GHz EVN image of J0849$+$1114 taken on 2019 Jan 22. The peak corresponds to the location of Galaxy 1. The peak brightness is $0.6\mathrm{\,mJy\,beam}^{-1}$, the contours are drawn at $(-2.4, 2.4, 5, 7.2) \times 64 \mathrm{\mu Jy\,beam}^{-1}$ ($1\sigma$ image noise level). The restoring beam is $4\mathrm{\,mas}\times2.6\mathrm{\,mas}$ (FWHM) with a major axis position angle $6\fdg8$ as shown in the lower left corner of the image. 
              }
         \label{fig:map}
   \end{figure}

To obtain the lowest noise-level image of J0849$+$1114, natural weighting was used and instead of several point-like {\sc clean} components, a single circular Gaussian brightness distribution model was fitted to the visibility data. The value of the reduced $\chi^2$, $1.4$, indicates an adequate fit to the data. The resulting image of J0849$+$1114 is shown in Fig. \ref{fig:map}. 

Due to the faintness of the target source, no self-calibration was attempted. The imperfect phase correction can result in a coherence loss which can lead to a flux density loss in the detected source. To estimate the amount of coherence loss, we turned to the candidate phase calibrator source, J0850$+$1108. We mapped the dataset obtained after fringe-fitting of J0850$+$1108 and the one obtained by phase solution transfer from the phase calibrator. In the latter case, the recovered flux density was $\sim 25$\,\% lower compared to fringe-fitting. This is in agreement with typical coherence loss values \citep[e.g.,][]{devos}.

\section{Results}

%

We detected a radio feature slightly extended to the north in J0849$+$1114. Using the {\sc aips} task {\sc maxfit}, we derived the position of the brightness peak, right ascension $08^\mathrm{h} 49^\mathrm{m} 05\fs 5181$, and declination $11\degr 14\arcmin 47\farcs 633$. The positional uncertainty of the phase calibrator source is $0.27$\,mas.\footnote{Global VLBI solution rfc\_2019a catalog, available at the astrogeo.org website, maintained by L. Petrov} By far the dominant contribution to the positional uncertainty ($\sim 3$\,mas) is caused by the angular separation between the target and the phase reference source \citep[e.g.,][]{chatterjee,rioja}. Thus, the derived coordinates are accurate within $\sim 3$\,mas.


The integrated flux density of the fitted Gaussian brightness distribution is $3.7\pm 0.1$\,mJy. After accounting for the coherence loss, the flux density of the radio feature is $S=5.0\pm 0.1$\,mJy. In the following, we use this flux density value. The full width at half maximum (FWHM) size of the Gaussian model component, $\theta=8.3 \pm 0.2$\,mas (corresponding to $\sim 12$\,pc projected size), is larger than the smallest resolvable size of this observation \citep[$1.3$\,mas, ][]{Kovalev2005}, therefore it can be used to derive the brightness temperature of the mas-scale radio emitting feature: 
\begin{equation}
T_\mathrm{B}= 1.22 \times 10^{12} \frac{S}{\theta^2\nu^2}(1+z) \mathrm{\,K}
\end{equation}
where $S$ is measured in Jy, $\theta$ in mas, $\nu$ is the observing frequency in GHz. The resulting brightness temperature, $3.3 \times 10^{7}$\,K, is two orders of magnitude higher than the brightness temperature limit measured above $1$\,GHz for star forming-galaxies \citep[$10^5$\,K, ][]{Condon1992}. 

Assuming a flat radio spectrum ($\alpha=0$), the $1.7$-GHz radio power is $6.7 \times 10^{22}\mathrm{\,W\,Hz}^{-1}$. (Owing to the small redshift of the source, the spectral index assumption does not significantly influence the obtained radio power.) According to \cite{Kewley2000} and \cite{Middelberg2011}, the radio power of an AGN core exceeds $2\times 10^{21}\mathrm{\,W\,Hz}^{-1}$. Few violent star-forming galaxies are known to have supernova complexes exceeding this value, e.g., Arp 299-A and Arp-220, with a $1.7$-GHz power of $4.5\times 10^{21}\mathrm{\,W\,Hz}^{-1}$ and $1.6\times 10^{22}\mathrm{\,W\,Hz}^{-1}$, respectively \citep[][and references therein]{alexandroff_2012}. However, the radio power of the detected feature J0849$+$1114 surpasses these values as well.

We did not detect any other mas-scale radio feature down to a $7\sigma$ noise-level of $0.45 \mathrm{\,mJy\,beam}^{-1}$ within a region of $8\arcsec \times 8\arcsec$ around the source that covers the cores of all galaxies in the interacting system. 
According to the EVN sensitivity calculator\footnote{\url {http://old.evlbi.org/cgi-bin/EVNcalc}}, bandwidth smearing and time-average smearing \citep{Wrobel1995} limit the undistorted field of view of our EVN observation to an area with radii $9\farcs9$ and $16\farcs7$, respectively. Thus, all three galaxy cores are within the observed field, where the smearing effects would not reduce significantly the response to a point source (the expected flux density loss is $<10\,\%$).

\section{Discussion}

The detected mas-scale radio feature (Fig. \ref{fig:map}) is positionally coincident with Galaxy 1 of \cite{Pfeifle2019}. Its high brightness temperature and 1.7-GHz radio power indicate that the radio emission originates from an AGN. Thus, we can confirm there is a radio-emitting AGN in Galaxy 1. 

Using the fundamental plane of black hole activity \citep{gultekin}, the black hole mass can be estimated from the $2-10$\,keV X-ray ($L_\mathrm{X}$), and the $5$-GHz radio luminosity. \cite{Pfeifle_new} gave $L_\mathrm{X}=(2.37 \pm 0.17) \times 10^{41} \textrm{\,erg\,s}^{-1}$ for the core of Galaxy 1 using a combination of {\it Chandra} observations conducted in 2013 and 2016. To estimate the $5$-GHz luminosity of the mas-scale radio core, we use the spectral index range found by \cite{Giroletti2009}, $-0.7 \le \alpha \le 0.1$, for the mas-scale cores of faint Seyfert galaxies. The obtained black hole mass estimates range between $\sim 4.6 \times 10^8 \textrm{\,M}_\odot$ (for $\alpha=-0.7$) and $\sim 2.9 \times 10^9 \textrm{\,M}_\odot$ (for $\alpha=0.1$). \cite{Pfeifle_new} reported the detection of broad Pa$\alpha$ emission lines in Galaxies 1 and 3. According to the relationship between the emission line and the black hole mass \citep{emBH}, the measured value implies a black hole mass of $3.2 \times 10^{7}\textrm{\,M}_\odot$ for Galaxy 1. Since the internal scatter of both relationships are $1$\,dex, and because of the non-simultaneous X-ray and radio observations the estimate from the fundamental plane can also be affected by brightness variability, we can regard the obtained black hole masses as not incompatible. These estimates all indicate that the measured X-ray, radio and Pa$\alpha$ emission line characteristics can be explained by an SMBH residing in the core of Galaxy 1.

The recovered flux density in our EVN observation, $S=5$\,mJy is much less than the one measured at a close frequency, $1.4$\,GHz within the FIRST survey \citep{first_cat}, $S_\mathrm{FIRST}=35.4\pm 0.1$\,mJy. While it could also be related to source variability, it is most likely that the missing $\sim 30$\,mJy flux density is in more extended structure for which our EVN observation is not sensitive. The largest recoverable feature in an interferometric observation is defined by the shortest baseline in the array \citep{Wrobel1995}. In our case, this is the Effelsberg$-$Westerbork baseline with $\sim 266$\,km. The corresponding largest recoverable size at $1.7$\,GHz is $\sim 70$\,mas, thus significant radio emission may be contained in features extending above this limit.  

The low resolution of FIRST ($\sim 5\arcsec$) does not allow us to relate the radio emission to any of the three interacting galaxies. The 1.4-GHz radio power corresponding to the flux density not detected by our EVN observation ($\sim 30$\,mJy) is $\sim 4 \times 10^{23}\mathrm{\,W Hz}^{-1}$. If we assume that this is caused only by star formation in the galaxies, it implies a summed star formation rate of $\sim 222 \mathrm{\,M}_\odot \mathrm{yr}^{-1}$ \citep{hopkins_sfr}. \cite{izotov_sfr}, who used multi-wavelength data to derive the spectral energy distribution and the star formation rate for a large sample of galaxies, obtained a star formation rate of $1.1\mathrm{\,M}_\odot \mathrm{yr}^{-1}$ for the galaxy group J0849$+$1114. \cite{Pfeifle_new} estimated an upper limit for the star formation rate assuming no AGN contribution in the system. The summed value for the three galaxies is $\sim 15 \mathrm{\,M}_\odot \mathrm{yr}^{-1}$. 
These numbers indicate that star formation alone cannot be responsible for this $30$\,mJy flux density, and AGN-related processes must dominate the $1.4$-GHz radio emission. The rms noise level of our map implies that there is no other compact radio-emitting AGN with $1.7$-GHz radio power larger than $\sim 6\times 10^{21}\mathrm{\,W Hz}^{-1}$ in a region encompassing the two other galaxies. However, the existence of an extremely low-power radio AGN cannot be ruled out. Even so, most of the AGN-related emission must be in large, at least hundred mas-scale lobe(s).

\cite{nature_cikk} report the results of a VLA A configuration observation of J0849$+$1114 at $9$\,GHz. They detected radio emission at the position of Galaxy 1 with a flux density of $18.65\pm 0.57$\,mJy and peak intensity of $5.7\mathrm{\,mJy\,beam}^{-1}$, and at the position of Galaxy 3 with a flux density of $2.28\pm 0.07$\,mJy and peak intensity of $0.5\mathrm{\,mJy\,beam}^{-1}$. 
Assuming $\alpha= 0$ for the compact, mas-scale core for the AGN in Galaxy 1, and subtracting $5$\,mJy from the flux density measured at $9$\,GHz, the emission at arcsec scales is expected to be $\sim 16$\,mJy. Then the spectral index of the arcsec-scale structure between $1.7$ and $9$\,GHz can be estimated, $\alpha \sim -0.4$. This value is close to the average spectral index, $-0.5$ derived for a sample of Seyfert galaxies using VLA observations by \cite{Ho2001}. Thus, the radio flux densities measured at high and low resolutions and at $1.7$ and $9$\,GHz can be explained with a compact core and extended emission region related to radio-emitting AGN. However, whether the radio emission detected at the position of Galaxy 3 by \cite{nature_cikk} is related to the AGN in that Seyfert galaxy, or if it is an extended jet-like feature of the radio AGN in Galaxy 1 cannot be ascertained. More sensitive VLBI (to reveal the faint radio-emitting AGN in Galaxy 3 if it exists) and/or intermediate resolution multi-frequency radio interferometric observations (to map the structure and spectral shape of the radio emission between Galaxy 1 and 3) could answer this question.


\section{Summary}
We conducted $1.7$-GHz high-resolution VLBI observation of the merging triple galaxy system, J0849$+$1114, which shows convincing multi-wavelength evidence for a triple AGN system \citep{Pfeifle_new, Pfeifle2019, nature_cikk}. We detected a high brightness temperature, compact radio feature in Galaxy 1, thus we can confirm the existence of a radio AGN there. Compared to lower resolution radio data, a significant amount of flux density remained undetected in our EVN observation. The amount of star formation derived for the system from independent methods cannot be responsible for most of this flux density. Thus, AGN-related emission, probably from $100$ mas-scale lobe-like feature(s) must contribute to it. Whether the large-scale emission is related to the EVN-detected compact AGN in Galaxy 1 or to a very low luminosity and thus by us undetected AGN in another nucleus is not clear. Nevertheless, the latter scenario is suggested by the $9$-GHz radio observation of \cite{nature_cikk} which showed arcsec-scale radio emission at the position of Galaxy 3. 


\begin{acknowledgements}
K.\'E.G. was supported by the J\'anos Bolyai Research Scholarship of the Hungarian Academy of Sciences and by the Ministry of Human Capacities within the framework of the \'UNKP (New National Excellence Program).
The European VLBI Network is a joint facility of independent European, African, Asian, and North American radio astronomy institutes. Scientific results from data presented in this publication are derived from the following EVN project code: RS09b. e-VLBI research infrastructure in Europe is supported by the European Union's Seventh Framework Programme (FP7/2007-2013) under grant agreement number RI-261525 NEXPReS. This publication has received funding from the European Union's Horizon 2020 research and innovation programme under grant agreement No. 730562 [RadioNet].
\end{acknowledgements}

%
%

\bibliographystyle{aa} 
\bibliography{ref} 

\end{document}